\documentclass[a4paper,fleqn,usenatbib]{mnras}
\usepackage{newtxtext,newtxmath}
\usepackage[T1]{fontenc}
\usepackage{ae,aecompl}
\usepackage{graphics}
\usepackage{graphicx}
\usepackage{amsmath}	
\usepackage{booktabs,caption}

\title[4U 1907+09]{A high-mass X-ray binary pulsar 4U 1907+09 with multiple absorption-line features in the spectrum}

\author[Tobrej et al.]{
Mohammed Tobrej,$^{1}$\thanks{tabrez.md565@gmail.com}
Binay Rai,$^{1}$\thanks{binayrai21@gmail.com}
Manoj Ghising,$^{1}$\thanks{manojghising26@gmail.com}
Ruchi Tamang,$^{1}$\thanks{ruchitamang76@gmail.com}
\newauthor
Bikash Chandra Paul$^{1}$\thanks{bcpaul@associates.iucaa.in}
\\
$^{1}$Department of Physics, North Bengal University, Siliguri, Darjeeling, WB, 734013, India
\\
}

\pubyear{2022}

\begin{document}
\label{firstpage}
\pagerange{\pageref{firstpage}--\pageref{lastpage}}
\maketitle

\begin{abstract}
 We report X-ray observations of the High Mass X-ray Binary (HMXB) pulsar 4U 1907+09. Spectral and Timing analysis of the source has been performed using NuSTAR observation. Timing analysis of the photon events revealed the coherent X-ray pulsation of the source with a pulse period of $442.92\;\pm\;0.03$ s. It is observed that the source is spinning down at a rate of $0.1971(4) s yr^{-1}$. The pulse profile is characterized by a decaying amplitude of the secondary peak and relative growth in the amplitude of the primary peak with an increase in energy. The broad-band spectral coverage of NuSTAR has been used to observe multiple absorption features in the X-ray continuum of the source. We confirm the presence of two prominent cyclotron absorption features  at $\sim 17$ keV and $\sim 38$ keV respectively. In addition, we have detected an absorption-line feature at $\sim 8$ keV, with an equivalent width of $\sim 1.3$ keV. The variation of the spectral parameters with pulse phase has been observed using phase-resolved spectroscopy and the relevant variabilities of the parameters have been discussed with the underlying physical implications. The continuum evolution and variations in spectral parameters have also been studied by time-resolved spectroscopy.  
\end{abstract}


%
%

\section{Introduction}
4U 1907+09 belongs to the class of the high-mass X-ray binaries (HMXBs) \citep{c}. The X-ray source was discovered in the Uhuru surveys \citep{a,b} and is known to be powered by wind accretion from a close blue supergiant companion star. The source was identified as a deep red early star with broad, strong $H_{\alpha}$ emission \citep{d}. According to \cite{d}, 4U 1907+09 is a binary system comprised of an OB supergiant star and a compact object, where the X-ray emission originates due to the accretion of the stellar wind onto the compact object. This hypothesis proposed by \cite{d} was validated by \cite{e} who revealed an orbital period of $\sim8.38$ days and the source was confirmed as a new member of a class of HMXBs (see,
e.g., \cite{f} for a review on these objects). Further, \cite{h} used optical spectroscopy, to show that the companion of the X-ray pulsar is a late O-type supergiant star \citep{i,j}. The categorization of the source was also refined by \cite{k} by observing an infrared spectrum of the source. A 437.5 s pulsation of the source was first detected by \cite{g}. The source is also known to exhibit occasional quasi-periodic oscillations (QPO) at a frequency of about 65 mHz \citep{H,m}. The source was initially found to exhibit a nearly constant spin-down rate \citep{n} but later observations indicated variability in the spin-down rate along with multiple torque reversals \citep{o,p}. The pulse period estimations made using INTEGRAL and RXTE observations have been found to exhibit short term fluctuations in pulse period over the long-term changes in the spin period rates which are consistent with the random walk model \citep{q}.

The X-ray continuum of the source reveals multiple prominent absorption features. Cyclotron Resonance Scattering Features (CRSFs) are distinctly observed in the present study along with an additional absorption-line feature that has not yet been studied in detail. Cyclotron absorption lines are significant characteristic features that have been observed in the spectrum of accreting X-ray pulsars. Such absorption features originate due to the resonant scattering of hard X-ray photons and electrons in quantized energy states \citep{r}. The energy levels (Landau levels), is equally spaced and are dependent on the magnetic field strength. The separation between Landau levels are equivalent to the centroid energy of the cyclotron. The cyclotron line energy is linked with the magnetic field strength by the relation,
\begin{equation}
\label{eqn:i}
$$E_{cyc}$ = 11.6 $\times\; B_{12} \times\; (1 + z)^{-1}$$(keV)
\end{equation}, where $ B_{12}$ is the magnetic field normalized in units of $10^{12}$ G and z ($\sim0.3$) represents the gravitational redshift in the scattering region for standard neutron star parameters. Hence, the detection of characteristic cyclotron lines provides a simple method for the direct estimation of the magnetic field of accreting X-ray pulsars. The characteristic cyclotron feature has been detected in several X-ray pulsars but there are relatively many accretion-powered pulsars in which such features have not been detected yet even with the surplus availability of data from various X-ray observatories.

4U 1907+09 is a well studied source and various observations related to the characteristic absorption feature have been reported in the literature. CRSF at 19 keV was first reported with Ginga \citep{Makishima and Mihara,Makishima} while the corresponding first harmonic was detected at 36 keV with BeppoSax observation \citep{Cusumano}. Further observations have been carried out with different instruments such as RXTE \citep{o, p}, INTEGRAL \citep{Hemphill}, Suzaku \citep{Rivers, Maitra and Paul} and AstroSat \citep{t}. In addition to the cyclotron absorption lines at $\sim 17$ keV and $\sim 38$ keV, an absorption feature at $\sim 8$ keV has been observed in the X-ray spectra of the source (Section \ref{4}). The centroid energy of this broad absorption line indicates an absorption related to the K-$\beta$ transition of the Fe XXV ion or an absorption line related to nickel . So, the origin of this absorption line remains suspicious.

A source distance of 5 kpc \citep{h} has been reported in the literature. However, the distance to the source estimated by Gaia EDR3 is about 1.9 kpc \citep{bailer}. Therefore, we have used a source distance of 1.9 kpc for estimating the X-ray luminosity. The flux in the (3-50) keV energy range is estimated to be $\sim6.5 \;\times\;10^{-10} erg\;cm^{-2}\;s^{-1}$ (Section \ref{section:4}) which corresponds to a luminosity of $\sim2.81\;\times\;10^{35} erg\;s^{-1}$ (Section \ref{section:4}) considering the source to be at a distance of 1.9 kpc. In this paper, we present detailed coverage of spectral and timing analysis of the source using NuSTAR observation. We report the detection of an additional absorption-line feature alongside two cyclotron lines in the X-ray spectra of the source. We have examined the reliability of the observational claims of the source designated by observation ID 30401018002 .

\section{Observation and Data reduction}
The source 4U 1907+09 was observed by Nuclear Spectroscopic Telescope Array (NuSTAR) observatory on August 01, 2018. Based on the ephemeris of \cite{p}, the NuSTAR observation under consideration covers the orbital phases (0.35-0.56). The data reduction has been carried out using HEASOFT v6.30 \footnote{\url{https://heasarc.gsfc.nasa.gov/docs/software/heasoft/download.html}} and CALDB version 20220525. NuSTAR is sensitive in the (3-79) keV energy range and is known to be the first hard X-ray focusing telescope. It is comprised of two independent co-aligned grazing incident telescopes that are similar but unidentical. Each telescope is assigned with its focal plane module FPMA and FPMB consisting of a pixelated solid state CdZnTe detector \citep{15}. Using the mission-specific  NUPIPELINE, clean event files were created for performing the analysis. The XSELECT tool was used for reading the obtained clean event files. The image was observed using astronomical imaging and data visualization application DS9\footnote{\url{https://sites.google.com/cfa.harvard.edu/saoimageds9}}. Spectra and light curves have been extracted by considering a circular region of $80^{\prime \prime}$ on detector 1 as the source region while backgrounds have been generated from an $80^{\prime \prime}$ region on detector 2 taken away from the source. The source and background files were used for obtaining the required light curve and the spectra by imposing the mission-specific task NUPRODUCTS. The background correction for the light curve was carried out using FTOOL LCMATH. The NuSTAR data has been barycentered to the solar system frame using the FTOOL BARYCORR. The spectra obtained were fitted in XSPEC version 12.12.1 \citep{s}. 
\begin{table}
\begin{center}
\begin{tabular}{ccccc}
\hline
Observation ID &  Observation Date (DD-MM-YYYY) &	Exposure (ks) \\	
\hline
30401018002	&	01-08-2018	&	78.86 \\

\hline
\end{tabular}

\caption{NuSTAR observation indicated by the observation ID along with the date of observation and exposure.}  
\end{center}
\end{table} 

\section{Timing Analysis}
\begin{figure}

\begin{center}
\includegraphics[angle=0,scale=0.35]{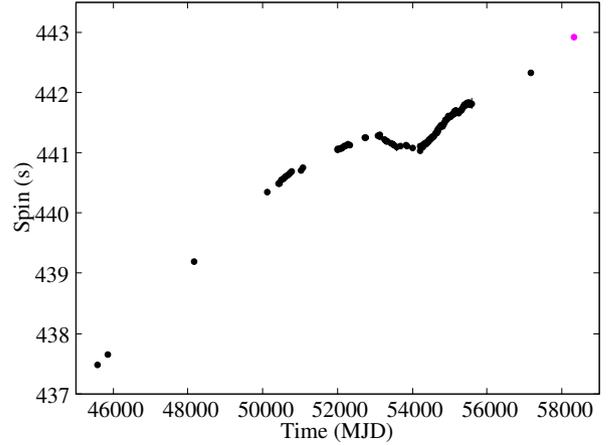}
\end{center}
\caption{Spin period history of 4U 1907+09. The points marked in black represent previously reported spin-periods while the point marked in magenta represents the spin-period corresponding to this work. }
\label{1}
\end{figure}
\begin{figure}

\begin{center}
\includegraphics[angle=0,scale=0.30]{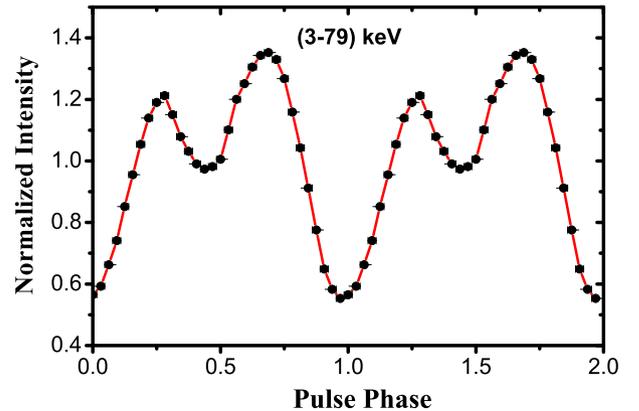}
\end{center}
\caption{Folded pulse profile of 4U 1907+09 in (3-79) keV energy range using 32 bins. The red line connects the phase points in linestep mode. The pulse profile has been normalized about the average count rate.}
\label{2}
\end{figure}

\begin{figure}

\begin{center}
\includegraphics[angle=0,scale=0.30]{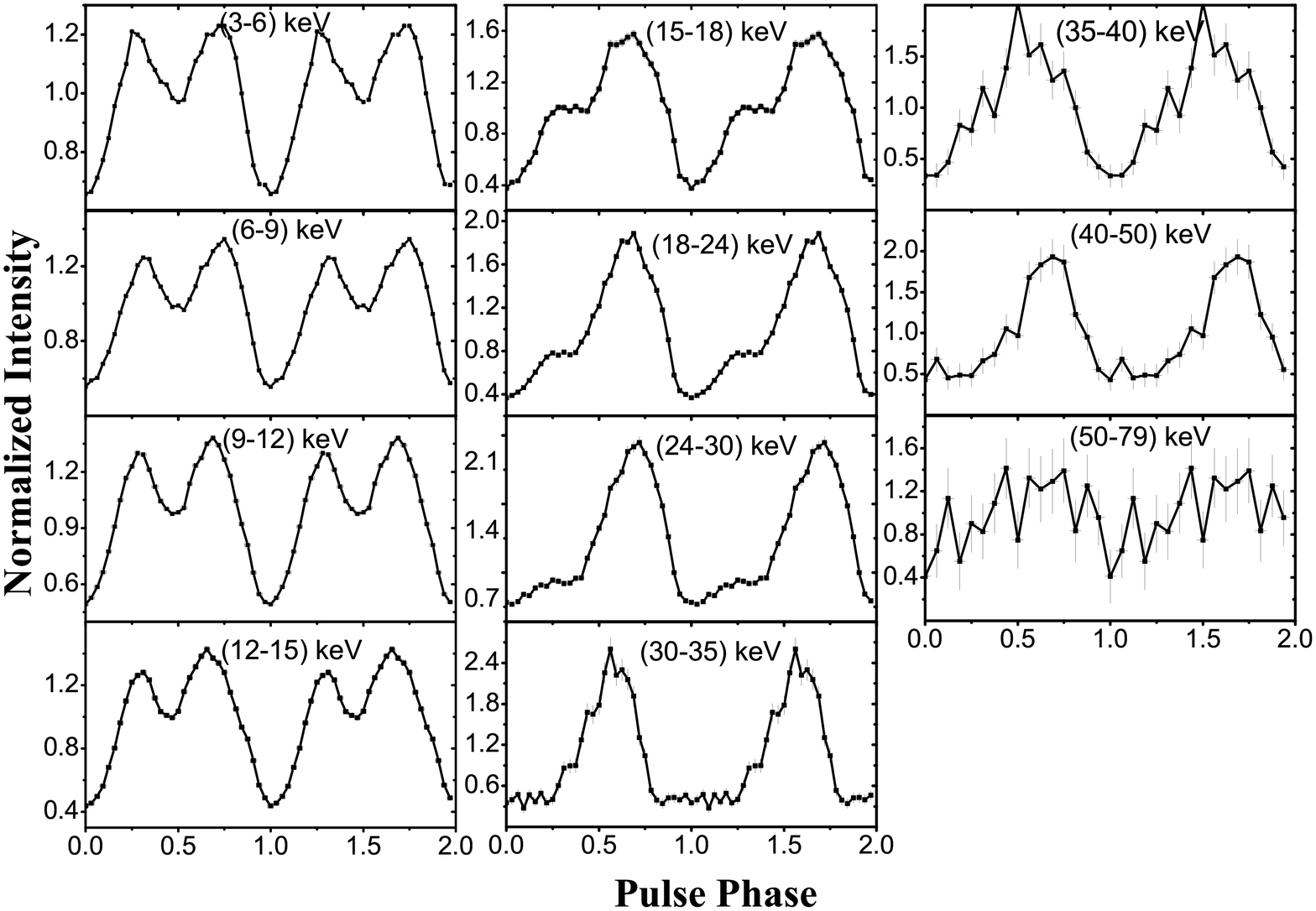}
\end{center}
\caption{Energy-resolved Pulse profiles coresponding to NuSTAR observation of 4U 1907+09.}
\label{3}
\end{figure}

\begin{figure}

\begin{center}
\includegraphics[angle=0,scale=0.35]{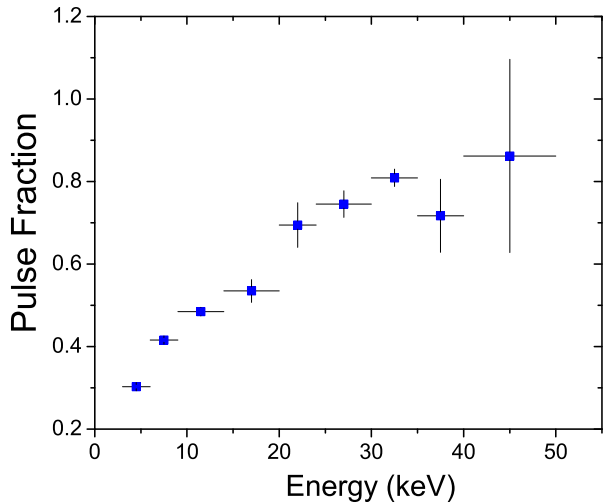}
\end{center}
\caption{ Variation of pulse fraction of the source with energy using NuSTAR observation.}
\label{4}
\end{figure}

We have considered NuSTAR light curves with a binning of 0.01 s for analyzing the temporal properties of the source. Light curves have been generated using FTOOL LCURVE. The approximate estimation of the pulse period was made using the Fast Fourier Transform (FFT) of the light curve. We performed the epoch-folding technique \citep{16,17} based on $\chi^{2}$ maximization for  measurement of the pulse period. Hence, we estimated the best period of the source to be $442.92\;\pm\;0.06$ s. We implemented the method described in \cite{18} for estimating the uncertainty in the measurement of pulsations. 1000 simulated light curves were generated and the respective pulsations for each curve were estimated. The standard deviation and the standard error was computed for the best estimation of the pulse period of the source. The tool EFSEARCH gives the best pulse period but does not generate the folded pulse profile of the source. Hence, the tool EFOLD is used to obtain the folded pulse profile of the source. The most recently reported pulse period estimation of the source is $442.33\;\pm\;0.07$ s \citep{t}. Therefore, it is apparent that the pulsar continues to spin down.

For demonstrating the torque reversal and the recent spin-down trend of 4U 1907 +09, we have used the pulse period history. The measurements of the pulse period revealed that the source exhibited a spin-down trend until 2003 \citep{H}. The spin-down rate was found to decrease significantly between 1998-2003 \citep{n}, and  a torque-reversal was observed in 2004 \citep{o}. Further observations by \cite{p,t} have revealed a spin-down trend of the source. Considering the spin value estimations from 2004 to 2018, it is observed that the source continues to spin down at a rate of $0.1971(4) s\; yr^{-1}$\;. Pulse-period history of the source presented in Figure \ref{1} has been plotted using the reports given by \cite{p}.

\subsection{Light curves, Pulse profiles and Pulse fraction}
The broadband coverage of NuSTAR in the energy range (3-79) keV permits us to explore the properties of the source.The pulse profile of the source in the (3-79) keV energy range is apparent to be dual-peaked in which the secondary peak is observed to exhibit a strong energy dependence. The secondary peak is characterized by a decaying amplitude along with a relative increase in the amplitude of the primary peak. For examining the emission pattern at different energies, pulse profiles of the source have been obtained by resolving the light curve in the energy range (3-79) keV into several energy bands of (3-6) keV, (6-9) keV, (9-12) keV, (12-15) keV, (15-18) keV, (18-24) keV, (24-30) keV, (30-35) keV, (35-40) keV, (40-50) keV \& (50-79) keV. The pulse profiles are folded with a chosen time zero-point $T_0$ (folding epoch) so that the minimum flux bin lies at phase point = 0.0. The pulse profiles are found to exhibit significant dependence on energy. The energy-resolved pulse profiles are observed to exhibit significant variability and are asymmetric in nature. The dual peaked structure exhibits a significant departure transforming to almost a single peaked structure at energies above 18 keV. The apparent single peaked profile is found to exhibit a distinct broadening above 35 keV. Comparing the strength of modulation is difficult along the different energy bands due to variations in the count rate. The significant evolution of the pulse profiles observed using NuSTAR data illustrates a changing activity in the accretion geometry of the pulsar. The folded pulse profile of the source in (3-79) keV energy range and the energy-resolved pulse profiles have been presented in Figure \ref{2} and Figure \ref{3} respectively.

The Pulse Fraction (PF) represents the relative amplitude of the emerging pulse profile. It relates to X-ray emission from the accretion column (pulsed emission) and other regions of the accretion flow or neutron star (NS) surface (unpulsed emission) \citep{w}. It is defined as, $PF=\;\frac{P_{max}-P_{min}}{P_{max}+P_{min}}$ where $P_{max}$ and $P_{min}$ represent maximum and minimum intensities of pulse profile respectively. The PF is found to follow an overall increasing trend with energy which is typically observed in most X-ray pulsars \citep{37}. However, the pulse fraction reveals a decreasing tendency around certain energies which is suggestive of absorption components revealed in the X-ray spectrum of the source. A decrease in the pulse fraction around the cyclotron line energy has been observed in the past for some X-ray pulsars \citep{TSY, 39, 37, tsy, lut}. The variation of Pulse Fraction with energy is presented in Figure \ref{4}.

\section{Spectral Analysis}
\label{section:4}
\begin{figure}
\begin{center}
\includegraphics[width=9cm, height=7cm, angle=-90]{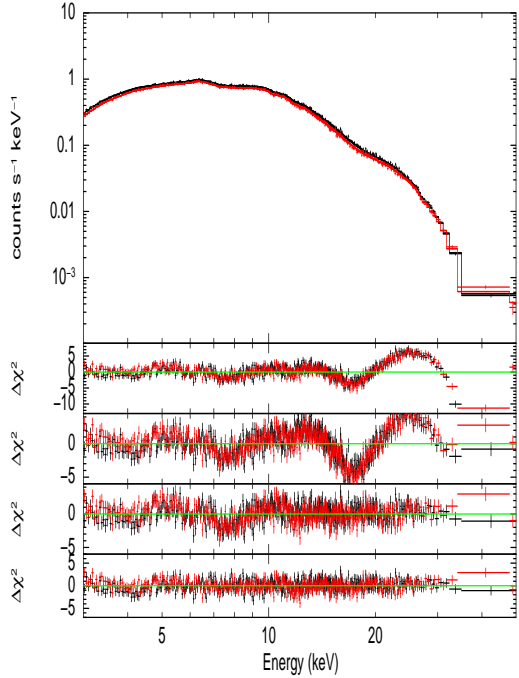}
\end{center}
\caption{The spectra corresponding to NuSTAR observation using continuum model I. The top panel represents the folded spectra  and the second panel (from top) represents the residuals without including GABS model. Panels below it represent the residuals after successive incorporation of GABS model for fitting the residuals at $\sim 38$ keV, $\sim 17$ keV and $\sim 8$ keV. Black \& red colours represent NuSTAR FPMA \& FPMB spectra respectively. The spectra has been rebinned for representative purpose.}
\label{5}
\end{figure}

The NuSTAR (FPMA \& FPMB) X-ray spectra of the source 4U 1907+09 were fitted in the broad-band (3-50) keV energy range and has been presented in Figure \ref{5}. This energy range was chosen due to background contamination above 50 keV. The FPMA \& FPMB data were grouped using the tool GRPPHA for obtaining a minimum of 20 counts per spectral bin. We ensured the relative normalization factors between the two modules  by constraining the constant factor associated with instrument FPMA as unity without imposing any constraint to the instrument FPMB such that the average count rate is comparable to that in FPMA. The constant factor corresponding to instrument FPMB  was found to be 0.994$\;\pm\;0.002$ which is relevant from the obtained statistical data.  This reveals an uncertainty of  $\sim 1\% $ which is in accordance with \cite{20}. The continuum emission of HMXB pulsars can be inferred to originate due to the Comptonization of soft X-rays in the plasma above the surface of the neutron star. We tried exploring several continuum models for obtaining the best fit spectral results. We modeled the absorption column density using the TBABS component with abundance from \citep{=}. The cross-section for the TABS component was chosen to be vern \citep{verner}. In the initial stage, the X-ray spectra were fitted by using the continuum model- CONSTANT*TBABS*(HIGHECUT*PO). The spectral fit obtained after imposing this continuum model was found to be associated with distinct negative and positive residuals. The positive residual was fitted by incorporating a GAUSSIAN component with the above continuum model revealing the presence of an iron emission line at 6.4 keV with an equivalent width of 0.23 keV. The fit statistics ($\chi^{2}$ per degrees of freedom) at this point was found to be (5701.20/1555) $\sim 3.66$. The X-ray spectrum was found to reveal three prominent absorption features which were appropriately fitted using the GAUSSIAN absorption components (GABS). As evident from the third panel of Figure \ref{5}, the absorption feature at $\sim38$ keV was fitted by incorporating a GABS component which improved the fit statistics to (5303.45/1563) $\sim 3.39$. The strength and width of the absorption line were found to be 16.15 keV and 5.11 keV respectively. We further introduced another GABS component to fit the absorption feature observed at $\sim17$ keV (Figure \ref{5}- fourth panel). The spectral fit obtained after including the two absorption components was improved significantly with a fit statistics of (2104.6/1551) $\sim 1.36$. However, an absorption feature at $\sim8$ keV was still distinctly observed in the source spectrum. This feature was also fitted by incorporating another GABS component, finally leading to a well-fitted spectrum of the source (Figure \ref{5} - bottom panel). Finally, the residuals observed in the X-ray spectra were well fitted by including all the above mentioned components. Hence, the final continuum model comprising three Gaussian absorption models was used for the best spectral fit of the source. The applied model estimated the best fit statistics to be (1806.20/1550) $\sim1.17$. The power-law photon index was found to be 0.79 with  cut-off energy $\sim13.66$ keV. The absorption features observed at $\sim17$ keV and $\sim38$ keV may be interpreted as cyclotron lines. The broad absorption feature observed at $\sim8$ keV may be an absorption feature related to  K-$\beta$ transition of the Fe XXV ion superimposed with Ni lines. This feature may also be a depression due to Ni $K_{\alpha}$ and Ni $K_{\beta}$ lines \citep{Furst}. The absorption column density (nH) was estimated to be $\sim(4.9\pm\;0.31)\times 10^{22}$ $cm^{-2}$. The estimated flux in the (3-50) keV energy range turns out to be $\sim\;6.5 \;\times\;10^{-10} erg\;cm^{-2}\;s^{-1}$ which corresponds to a luminosity of $\sim\;2.81\;\times\;10^{35} erg\;s^{-1}$ considering the source to be at a distance of 1.9 kpc \citep{bailer}.

To examine the significance of the observed spectral features, we also used the bulk and thermal Comptonization model (also known as Becker \& Wolff (BW) model) \citep{BeckerWolff} which fitted the spectra reasonably well indicating that the features are model-independent. This model is used to explain the observed spectrum of accreting X-ray pulsars and is based on the thermal and bulk comptonization of the seed photons emitted due to bremsstrahlung, cyclotron, and blackbody processes occuring in the accreting plasma. This model has been successfully employed to study the broadband X-ray spectra of X-ray pulsars like 4U 0115+63, 4U 1626-67, Her X-1, EXO 2030+375 , and LMC X-4 \citep{39, dai, wolff, epili, ambrosi}. \cite{BeckerWolffa, BeckerWolffb} explained the spectrum of various sources in a narrow energy range using this model. The Comptonization of the seed photons leads to a spectrum which can be explained by a power-law continuum with an exponential cut-off at high energies. The broadband spectrum of 4U 1907+09 was fitted by BW model along with the Gaussian emission component and three GABS components. The GABS components were employed to fit the three prominent absorption features in the X-ray spectra of the source. The radius and mass of the neutron star in BW model were fixed to the canonical values while the distance parameter was fixed at 5 kpc. Constraining the radius of accretion column $r_{0}$ was found to be difficult leading to insensitivity in the fitting. The five free parameters of the model are the diffusion parameter $\xi$, the ratio of bulk and thermal Comptonization $\delta_{0}$, mass accretion rate $\dot{M}$, magnetic field $B$, and electron temperature $T_{0}$. The blackbody, cyclotron and bremsstrahlung normalizations were also fixed to specific values according to the user manual of BW model \footnote{\url{https://www.isdc.unige.ch/~ferrigno/images/Documents/BW_distribution/BW_cookbook.pdf}}. The parameters corresponding to the absorption features are found to be consistent with the results obtained using other phenomenological models. The magnetic field estimated using BW model is found to be $\sim$ 2.12 $\times$10$^{12}$ G. Also, the magnetic field estimated using equation \ref{eqn:i} by considering the fundamental cyclotron line as 17.87 keV (Table \ref{2}) is $\sim$ 2.01 $\times$10$^{12}$ G. Therefore, it is evident that the magnetic field strength estimated using the BW model is consistent with the estimated field strength using the cyclotron line energy. The ratio of bulk to thermal Comptonization is estimated to be $\sim$ 0.24. The diffusion parameter is found to be $\sim$ 15.37. Similarly, the electron temperature is found to be 4.33 keV along with an estimated mass acretion rate of $\sim$ 1$\times$10$^{17}$ gs$^{-1}$. The $\chi^{2}$ of the overall fitting is found to be 1848.07 for 1549 degrees of freedom.

Further, a third continuum model (CUTOFFPL) was also used for spectral fitting. The fit parameters corresponding to this model were also found to be consistent with an estimated fit statistics of $\sim1.18$. The spectral fit parameters corresponding to the three continuum models have been presented in Table \ref{2}. 

To assess the statistical significance of the absorption features, we have performed 1000 Monte-Carlo simulations using the XSPEC script \textit{simftest}. For the continuum model consisting of (HIGHECUT*PO) along with a Gaussian emission component, the X-ray spectra revealed some negative residuals. We have employed a widely known absorption component GABS to explore the  presence of the absorption lines. The analysis gives the value of the statistical difference between the models, with and without specific GABS components. The large difference ($\Delta\chi^{2}$) of $\chi^{2}$  between the observed spectral fit and maximum value from the simulated data-sets for the given spectral features confirm the significance of the features in the spectra. The value of $\Delta\chi^{2}$ obtained by simulations is tremendously low in comparison to $\Delta\chi^{2}$ obtained using the actual data corresponding to the absorption features at $\sim 8$ keV, $\sim 17$ keV and $\sim 38$ keV respectively. Hence, a probabilistic origin of the absorption lines can be neglected. Also, there are no absorption features related to the NUSTAR detectors at around 8 keV. However, at around 10 keV, the tungsten (W) L-edge is known to cause residuals \citep{E}. Therefore, we can safely rule out an instrumental origin of the absorption line.  

\begin{table*}
\begin{center}
\begin{tabular}{cccc}
\hline										
Parameters		&		MODEL I (HIGHECUT) 	&		MODEL II (BW)	&		MODEL II (CUTOFFPL)		\\
\hline														
$C_{FPMA}$		&		1(fixed)	&		1(fixed)	&		1(fixed)		\\
$C_{FPMB}$		&		0.994$\pm$0.002	&		0.998$\pm$0.003	&		0.995$\pm$0.002		\\
$nH \;(cm^{-2})$		&		4.9$\pm$0.32	&		5.86$\pm$0.25	&		4.41$\pm$0.30		\\
$\xi$ 		&		-	&		15.37$\pm$0.58	&		-		\\
$\delta$		&		-	&		0.24$\pm$0.06	&		-		\\
$B$ ($\times 10^ {12} G$)		&		-		&		2.12$\pm$0.03		&		-		\\
$\dot{M}$ ($10^{17}gs^{-1}$)		&		-		&		1.01$\pm$0.08		&		-		\\
$T_{e}$ (keV)		&		-		&		3.19$\pm$0.008		&		-		\\
$r_{0}$ (m)		&		-	&		10(fixed)	&		-		\\
$\Gamma$		&		0.79$\pm$0.04	&		-	&		-0.42$\pm$0.04		\\
$E_{cut}$ (keV)		&		13.66$\pm$0.07	&		-	&		13.75$\pm$0.07		\\
$E_{fold}$ (keV)		&		7.95$\pm$0.28	&		-	&		-		\\
$E_{Fe}$ (keV)		&		6.4$\pm$0.005	&		6.37$\pm$0.004	&		6.41$\pm$0.002		\\
$\sigma_{Fe}$ (keV)		&		0.23$\pm$0.03	&		0.27$\pm$0.03	&		0.29$\pm$0.03		\\
$E_{gabs1}$		&		7.97$\pm$0.07	&		8.01$\pm$0.05	&		7.83$\pm$0.07		\\
$Sigma_{gabs1}$		&		1.26$\pm$0.11	&		2.14$\pm$0.19	&		1.61$\pm$0.08		\\
$D_{gabs1}$		&		0.33$\pm$0.07	&		1.16$\pm$0.04	&		0.76$\pm$0.07		\\
$E_{gabs2}$		&		17.29$\pm$0.009	&		17.87$\pm$0.007	&		17.99$\pm$0.005		\\
$Sigma_{gabs2}$		&		3.67$\pm$0.27	&		4.61$\pm$0.06	&		2.56$\pm$0.08		\\
$D_{gabs2}$		&		5.02$\pm$0.52	&		9.20$\pm$0.09	&		2.74$\pm$0.11		\\
$E_{gabs3}$		&		38.14$\pm$0.037	&		37.82$\pm$0.56	&		37.97$\pm$0.27		\\
$Sigma_{gabs3}$		&		5.11$\pm$0.32	&		7.47$\pm$0.42	&		4.38$\pm$0.21		\\
$D_{gabs3}$		&		16.15$\pm$1.34	&		19.15$\pm$0.34	&		12.17$\pm$0.81		\\

Flux($\times   10^{-10}\;erg\;cm^{-2}\;s^{-1}$)		&		6.5$\pm$0.06	&		6.41$\pm$0.07	&		6.48$\pm$0.06		\\
$\chi^{2}_{\nu}$		&		1.17	&		1.18	&		1.18		\\

 \hline
 \end{tabular}
 \caption{\label{tab:2} The fit parameters of 4U 1907+09 using three continuum models represented by MODEL I, MODEL II and  MODEL III respectively. Flux was calculated within energy range (3-50) keV for the NuSTAR observation. The absorption column density (nH) is expressed in units of $10^{22} cm^{-2}$. The fit statistics $\chi_{\nu}^{2}$  represents reduced $\chi^{2}$ ($\chi^{2}$ per degrees of freedom). D represents the strength of absorption lines. Errors quoted are within 1$\sigma$ confidence interval. The parameters corresponding to the BW model are represented by $\xi$ (diffusion parameter), $\delta$ (ratio of bulk to thermal Comptonization), B (magnetic field) $T_{e}$(electron temperature), $r_{0}$ (column radius), and $\dot{M}$ (mass accretion rate).} 
 \label{2}
  \end{center}
 \end{table*}

\subsection{Phase-Resolved Spectral Analysis}
The phase-resolved spectral analysis is employed to analyze the anisotropic properties of the X-ray emitted by the pulsar around its rotational phase. The significant energy dependence of the pulse profiles is evident from the energy-resolved profiles represented in Figure \ref{3} that reveals a highly significant variation in the morphology of the pulse profiles. The profile is found to transform from a dual-peaked structure to almost a single-peaked structure above 18 keV. Hence, it would be interesting to examine the dependence of the spectral parameters on changing viewing angle of the neutron star. We have divided the estimated pulse period of the source into 10 segments and generated the spectra corresponding to each segment. 
\begin{figure}

\includegraphics[angle=270,scale=0.33]{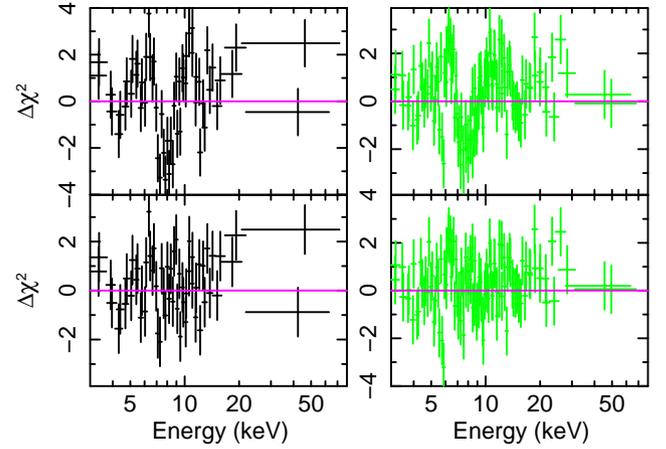}

\caption{Residuals corresponding to phase bins 0.45 (left) and 0.95 (right) where the absorption feature at around 8 keV is observed. The above figure represents the combined residuals of FPMA and FPMB. The spectra has been rebinned for representative purpose.}
\label{6}
\end{figure}

\begin{figure}

\includegraphics[angle=0,scale=0.3]{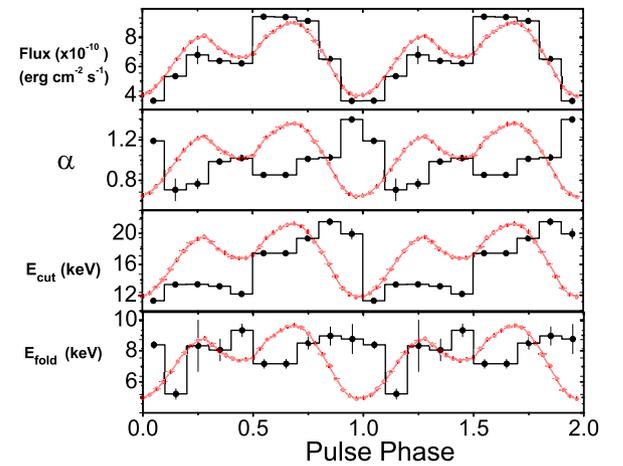}

\caption{Variation of spectral parameters: Flux, Photon Index ($\alpha$), Cut-off energy ($ E_{cut}$) and folding energy ($ E_{fold}$) with pulse phase for NuSTAR observation of 4U 1907+09.}
\label{7}
\end{figure}

\begin{figure}

\includegraphics[angle=0,scale=0.3]{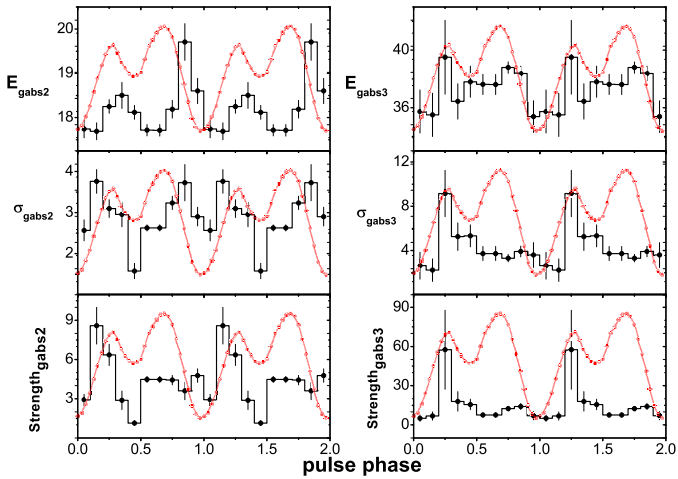}

\caption{Variation of cyclotron line parameters with pulse phase. The subscripts gabs2 and gabs3 represent the 17 keV and 38 keV absorption features respectively.}
\label{8}
\end{figure}

The spectra corresponding to the phase bins have been well fitted using MODEL I. All the spectral parameters were found to exhibit significant variabilities with the pulsed phase. The pulse phase variation of the spectral parameters and the CRSF parameters obtained using the above continuum model have been presented in Figure \ref{7} and Figure \ref{8} respectively along with the corresponding error bars. The flux has been estimated in the (3-50) keV energy range corresponding to each phase bin. The morphology of flux variation is presented in the top panel of Figure \ref{7} and is comparable to the nature of the pulse profile of the source. The power law photon index resembles a double-peaked profile with a phase shift relative to the pulse profile. The photon index is found to decrease near the peak emission of the profile suggesting a harder spectrum at that phase. Similarly, the photon index is found to be high near the minimum of the profile, which is suggestive of relatively soft spectra. This feature is in accordance with the increasing pulse fraction with energy. The photon index is found to range between 0.7 to 1.4 along the phase bins. The $E_{cut}$ and $E_{fold}$ parameters are also found to exhibit some variation with the pulse phase. The $E_{cut}$ parameter is found to range between (11.3-21.5) keV while the $E_{fold}$ parameter ranges in the limit (5.21-9.32) keV. The cut-off energy ($E_{cut}$) is found to attain maxima near the end of the primary peak in the pulse profile. 

The CRSF parameters are found to exhibit significant variations relative to the pulse phase. The centroid energy corresponding to the 17 keV cyclotron line is found to deviate by about 14\% while that of the 38 keV cyclotron line is found to deviate by about 6\% relative to the phase-averaged value. The centroid energy of the 17 keV cyclotron line attains a peak value near the maxima of the primary peak (phase interval (0.8-0.9)) of the pulse profile while the maxima corresponding to the 38 keV cyclotron line energy is observed in the phase interval (0.4-0.5) which is coincident with the secondary peak of the pulse profile. The width and strength of the two CRSFs are found to follow a complex variability with the pulse phase. The width of the two cyclotron lines ranges in the limit (1.58-3.72) keV and (2.23-9.15) keV respectively. Similarly, the strength of these lines is found to range in the limit (1.14-8.59) keV and (5.1-57.5) keV respectively. The strength corresponding to the two lines is found to peak near the maxima of the secondary peak of the pulse profile. The strength of the 38 keV  cyclotron line is nearly constant except in the phase interval (0.2-0.3) where an abrupt increase in the strength is observed. The 8 keV absorption feature was prominently observed along the phase bins 0.35, 0.45, 0.55, 0.65 and 0.95. The centroid energy corresponding to this feature is found to deviate by about 5\% relative to the phase-averaged value. The residuals corresponding to phase bins 0.45 and 0.95 have been presented in Figure \ref{6}. For phase bin 0.35, the fit statistics was found to improve from $ 1.16 $ to $1.02 $ on incorporating the GABS model. Also, the fit statistics was found to improve from $1.19 $ to  $1.06 $ in the phase bin 0.45. Similarly, the fit statistics in the phase bins 0.55, 0.65, and 0.95 were found to improve from $1.19$ to $1.13 $, $1.15 $ to $1.09 $ and $1.10 $ to $1.03 $ respectively by including the GABS model to fit the absorption feature.
\subsection{Time Resolved Spectral Analysis}
\begin{figure}

\includegraphics[angle=0,scale=0.31]{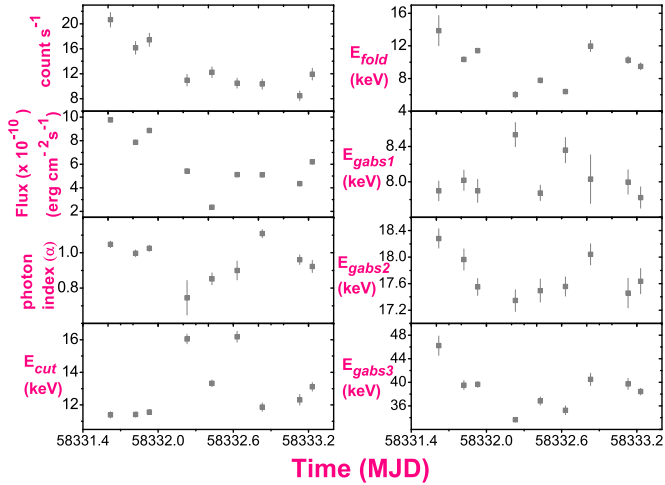}

\caption{Variation of spectral parameters: Count rate, Flux, Photon Index ($\alpha$), Cut-off energy ($ E_{cut}$), folding energy ($ E_{fold}$) and absorption line energies with time for NuSTAR observation of 4U 1907+09.}
\label{9}
\end{figure}

\begin{figure}
\begin{center}
\includegraphics[angle=0,scale=0.35]{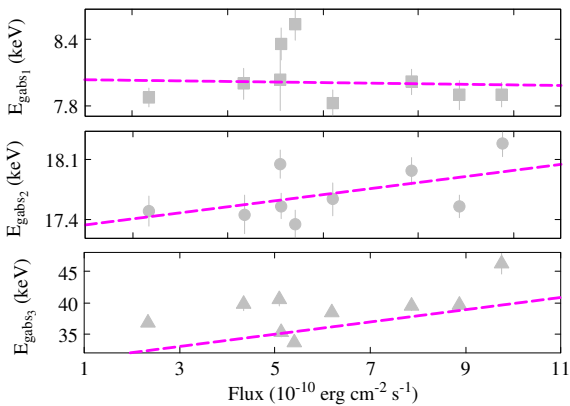}
\caption{Variation of centroid energies of 8 keV, 18 keV and 38 keV absorption features with flux. The dashed line represents the best fitted line.}
\label{10}
\end{center}
\end{figure}

The count rate of the source is found to follow an overall decaying trend with time (Figure \ref{9}- top panel (left)). Therefore, it would be interesting to examine the continuum evolution and variations in spectral parameters at different times. The source light curve was divided into nine segments and the corresponding spectra were generated by using good time interval (GTI) files for each segment. We have used model I for the best spectral fit. The source flux was found to range in the limit $\sim
(2.34 \;\times\;10^{-10}$ to $9.76 \;\times\;10^{-10})$ erg cm$^{-2}$ s$^{-1}$ revealing a significant variation along the nine segments of the light curve. Three distinct absorption features were observed in the X-ray spectra corresponding to all nine segments. The variations in the spectral parameters with time have been presented in Figure \ref{9}. The count rate and flux of the source were found to follow a decaying trend with time. The centroid energies corresponding to the absorption features were found to evolve with time. The other spectral parameters were found to vary moderately with time. Flux dependence of centroid energies corresponding to the absorption lines is presented in Figure \ref{10}. CRSFs at 17 keV and 38 keV seem to follow a similar trend, while the 8 keV feature follows a different trend with flux. The line energies corresponding to the 8 keV absorption feature were found to be weakly correlated with flux (correlation coefficient $\sim 0.2$). However, the line energies correponding to the 17 keV and 38 keV absorption features were found to be moderately correlated with flux characterized by a correlation coefficient of 0.58 and 0.62 respectively.   
\section{Discussion \& Conclusion}
The available NuSTAR data observed on August 01, 2018 has been used to analyze the spectral and temporal properties of this HMXB source. Timing analysis of the photon events detected the coherent X-ray pulsation of the source with a pulse period estimated to be  $ 442.92\;\pm\;0.03$ s. From 1983, when the pulsation of the source was discovered \citep{g}, the period had a steady spin-down rate of $0.225 \; s\; yr^{-1}$, with the spin period ranging from 437.5 s to 440.3 s \citep{H}. A deviation from the spin-down rate was reported by \citep{n}, revealing a lower spin-down rate between 1998 and 2003. In 2002, the spin-down rate was almost half the long-term value ($0.115\; s\;yr^{-1}$). Between 2004 and 2005, a complete torque reversal was observed \citep{o} with a maximum period of 441.3 s in 2004 April subsequently followed by spin up trend at a rate of - $0.158\;\pm\;0.007\; s\; yr^{-1}$.  A second torque reversal and a new spin-down rate of $0.220 \;s\; yr^{-1}$, consistent with the rate before 1998 was reported by \cite{p} using Rossi X-Ray Timing Explorer (RXTE)-PCA data between 2007 and 2008. The most recently reported pulse period of the source corresponding to the 2017 observations is $442.33\;\pm\;0.07$ s \citep{t} revealing that the source has been spinning down. It is apparent from the present study that the pulsar continues to spin down at a rate of $ 0.1971(4)\; s\; yr^{-1}$.
The source may be spinning down due to magnetic dipole radiation. Considering the previously reported flux measurements and spin trends observed by \cite{n}, \cite{o}, \cite{p} and \cite{t}, alongside our present estimation, it is observed that the X-ray flux does not exhibit clear evidence of a proper correlation between X-ray fluxes and the relevant changes in spin-down/ spin-up rates. Therefore, the spin period evolution of the source may be explored in the future using more detailed theoretical studies.
 
Pulse profiles of many pulsars have been found to transform from multi-peaked structure to a single peak structure with increasing energy \citep{19}. These types of transformation are typically observed mostly in bright sources \citep{37}. The pulse profiles of the source 4U 1907+09 are found to exhibit a strong dependence on energy. At lower energies, the profiles resemble a dual-peaked structure with a gradually decaying amplitude of the secondary peak. The transformation of the pulse profiles may be associated with a transition in the accretion regime of the pulsar. Furthermore, the pulse profiles are found to be asymmetric which is typical in X-ray pulsars. The asymmetric nature of the pulse profiles has been justified by several theoretical models. As per the reports of \citep{103,104,105,106}, the distorted magnetic dipole field where the  magnetic poles are not exactly opposite to one another has been quoted as a possible reason for the asymmetry of pulse profiles. An asymmetric accretion stream has also been marked as a probable reason for the asymmetric structure of pulse profiles \citep{y,101,z}. As the X-ray luminosity of the source turns out to be on the lower side, a pencil-beam pattern is expected in 4U 1907+09. The two peaks observed in the pulse profile arise from the magnetic poles. However, the primary peak and secondary peak are not discretely set apart by a phase of 0.5, which is suggestive of either a non-dipole field geometry of the neutron star or a complex beaming pattern from at least one of the poles. A complex beaming pattern resembles a mixture of pencil and fan beam accretion geometry. The contrasting energy dependence of the primary peak and secondary peak reflects a strong disparity in the energy dependence of the beaming pattern in the magnetic poles of the source.
 
The PF which constitutes the fraction of photons contributing to the modulated part of the flux resembles an interesting feature of the source. The variation of PF with energy is consistent with the presence of characteristic absorption observed in the energy spectrum of the source. The PF follows an overall increasing trend with energy but is characterized by dips in the form of local minima around certain energies which accounts for the prominent absorption features relevant in the source spectra. The non-monotonic dependence of the PF on energy about the cyclotron feature has been observed in various X-ray pulsars \citep{38,37,39}.
  
The NuSTAR spectrum has been well fitted by using three continuum models. The corresponding fit parameters have been presented in Table 2. In addition to an iron emission line at $\sim 6.4$ keV, multiple absorption features  are prominently observed in the X-ray continuum. The physical BW model reveals that the thermal Comptonization dominates the bulk Comptonization. The domination of thermal Comptonization over bulk Comptonization in the low luminosity state was observed in EXO 2030+375 by \cite{epili}. According to \cite{doro}, the two components in the broadband spectrum may be a result of bulk and thermal Comptonization close to the surface of the neutron star which can be suitably described using semi-phenomenological models. Cyclotron lines are observed at $\sim 17$ keV and $\sim 38$ keV respectively. After the detection of the first harmonic in 1998 \citep{Cusumano}, this feature has not been studied efficiently in the literature. Therefore, the detection of this rarely studied absorption feature at $\sim 38$ keV is significant. Interestingly, one more absorption feature at $\sim 8$ keV with an equivalent width of $\sim 1.3$ keV is distinctly observed in the spectra. Interpreting this feature as CRSF, the magnetic field strength is estimated to be $\sim 8.9 \times 10^{11}G$. We verified the estimated magnetic field using QPO calculations (e.g.,\cite{Raichur and Paul, Ketan, Jam}). Considering a QPO frequency of 65 mHz \citep{H,m}, the corresponding magnetic field strength is estimated to be  $\sim 4 \times 10^{12}G$ which is inconsistent with the value estimated by interpreting the 8 keV feature as CRSF. The centroid energies of Ni $K_{\alpha}$ and Ni $K_{\beta}$ lines are 7.5 keV and 8.26 keV respectively \citep{Sako, Palmeri}. Prominent detections of nickel have been observed only in a few astrophysical X-ray spectra which may be due to limited counting statistics at energies above 7 keV, where the nickel lines occur. Also, the poor spectral resolution in this band, makes nickel $K_{\alpha}$ and iron $K_{\beta}$ lines indistinguishable. Therefore, this broad absorption feature may be related to  K-$\beta$ transition of the Fe XXV ion superimposed with Ni lines. It may also be a depression due to Ni $K_{\alpha}$ and Ni $K_{\beta}$ lines \citep{Furst}. A similar study of the source was carried out by \cite{t} with AstroSat observation of June, 2017 which confirmed the presence of a cyclotron line at $\sim 18.5$ keV. However, the presence of other absorption features was not revealed in their study. 

The presence of distinct spectral features has also been examined by phase-resolved spectroscopy, as various sources are known in which the cyclotron line or its higher harmonics have been detected only at specific phases of rotation \citep{I}. The strength of the cyclotron line is found to depend strongly on the pulse phase for many sources. The fundamental line is sometimes observed only at certain pulse phases in sources like Vela X-1 \citep{+}, or KS 1947+300 \citep{-}. This suggests that the contributions of the two accretion columns are variable and/or that the emission pattern during the relevant pulse is such that the CRSF is very shallow or filled by spawned photons \citep{*}. According to the latter model, the harmonic line typically exhibits a comparatively less depth variability with the pulse phase. The pulse phase resolved spectroscopy with 10 independent phase bins reflect significant variability of the spectral parameters. The absorption feature at $\sim 8$ keV was observed in 5 phases and was not prominent in some phases. An absence or non-detection of the absorption line in some pulse phases may be a consequence of a large gradient in the strength of the magnetic field over the visible column height or latitudes on the surface of the neutron star \citep{99,x,C}. The other two absorption features at $\sim 17$ keV and $\sim 38$ keV have been distinctly observed in all the phase bins. The centroid energy corresponding to the 17 keV cyclotron line is found to deviate by about 14\% while that of the 38 keV cyclotron line is found to deviate by about 6\% relative to the corresponding phase-averaged values which is lower in comparison to sources like Vela X-1, 1A 1118-61, and XTE J1946+274 \citep{100}. The strength of the two absorption lines are found to attain peak values near the maxima of the secondary peak of the pulse profile which may be due to the difference in beaming pattern in the two magnetic poles or a possible divergence from dipole geometry of the neutron star's magnetic field. The power-law photon index was found to exhibit significant variability with the pulse phase of the system. As evident from Figure \ref{7} (Second panel), the photon index is harder when the source flux is high, while it is relatively softer when the source flux is low. The optical depth increases with the increase in luminosity leading to an increase in the hardness of photons which is in accordance with hardening (decrease in photon index) of the spectrum with increase in the flux \citep{40}. The average flux of the source in (3-50) keV energy range was found to be $\sim 6.5\;\times\;10^{-10}\;erg\;cm^{-2}\;s^{-1}$ which represents a luminosity of $\sim 2.81 \;\times\;10^{35}\;erg\;s^{-1}$ assuming a distance of 1.9 kpc.

For exploring the spectral evolution in the pulsar, we have performed time-resolved spectroscopy. The continuum is found to evolve with flux. The multiple absorption line features observed in the X-ray spectra have been found to vary with flux. The line energies have been found to to be positively correlated with flux which is typically observed in low to medium luminosity accreting binary pulsars. Based upon our results, a transition in the accretion regimes and the corresponding spectral variabilities or changes in emission geometry can be explored by future observations at different luminosities.

\section*{Acknowledgements}
This work has been done by using publicly available data provided by NASA HEASARC data archive. The NuSTAR data has been provided by NASA High Energy Astrophysics Science Archive Research Center (HEASARC), Goddard Space Flight Center. We are grateful to IUCAA Centre for Astronomy Research and Development (ICARD), Department of Physics, NBU, for providing research facilities. BCP would like to acknowledge DST-SERB for providing research grant. The authors are grateful to the anonymous reviewer for his/her valuable comments for enhancing the quality of the manuscript.

\section*{Data availability}
 
The observational data used in this work can be accessed from the HEASARC data archive and is publicly available for analysis.









\bsp	
\label{lastpage}
\end{document}